\documentclass[1p,numbers]{elsarticle}

\usepackage[english]{babel}
\usepackage[utf8x]{inputenc}
\usepackage{titlesec, amsmath,stackengine,graphicx,subfigure,tabularx,arydshln,url,stfloats}
\stackMath

\usepackage[tableposition=top]{caption}
\usepackage[colorinlistoftodos]{todonotes}
\usepackage[colorlinks=true, allcolors=blue]{hyperref}
\usepackage[left]{lineno}
\biboptions{sort&compress}
\title{No correlation between Solar flares and the decay rate of several $\beta$-decaying isotopes}

\begin{document}
\renewcommand\arraystretch{1.2}
\author[nikhef]{J.R.~Angevaare}
\author[uzh]{L.~Baudis}
\author[nikhef]{P.A.~Breur\corref{cor}}
\ead{sanderb@nikhef.nl}
\author[uzh]{A.~Brown}
\author[nikhef]{A.P.~Colijn}  
\author[purdue]{R.F.~Lang}
\author[cbpf]{A.~Massafferri}
\author[nikhef]{J.C.P.Y.~Nobelen\corref{cor}}
\ead{jaspern@nikhef.nl}
\author[cbpf]{R.~Perci}
\author[purdue]{C.~Reuter} 
\author[freib]{M.~Schumann}
\cortext[cor]{Corresponding Authors}
\address[nikhef]{Nikhef and the University of Amsterdam, Science Park, 1098 XG Amsterdam, the Netherlands}
\address[uzh]{Physik-Institut, Universit\"{a}t Z\"{u}rich, 8057 Zurich, Switzerland}
\address[purdue]{Department of Physics and Astronomy, Purdue University, West Lafayette, IN 47907, USA}
\address[cbpf]{Centro Brasileiro de Pesquisas F\'{i}sicas - COHEP, R. Dr. Xavier Sigaud, 150 - Urca, Rio de Janeiro, Brazil}
\address[freib]{Physikalisches Institut, Universit\"{a}t Freiburg, 79104 Freiburg, Germany}
\begin{abstract}
	We report on finding no correlation between the two strongest observed Solar flares in September 2017 and the decay rates of $^{60}$Co, $^{44}$Ti and $^{137}$Cs sources, which are continuously measured by two independent NaI(Tl) detector setups. We test for variations in the number of observed counts with respect to the number of expected counts over multiple periods with timescales varying from 1 to 109 hours around the Solar flare. No excess or deficit exceeds the 2$\sigma$ global significance. We set a conservative lower limit on the decay rate deviation over an 84 h period around the two correlated Solar flares in September 2017 to 0.062\% with 2$\sigma$ confidence. A fractional change of $~0.1\%$ in the decay rate of $^{54}$Mn over a period of 84 h was claimed with 7$\sigma$ significance during multiple Solar flares in December 2006. We exclude such an effect at 4.7$\sigma$ significance. 
\end{abstract}
	\bibliographystyle{abbrv}
\maketitle
\section{Introduction}
In the past decade, various claims~\cite{Jenkins2009,Jenkins2008,O'Keefe:2012,Jenkins2012B} have been made that the decay constants of certain isotopes' $\beta$-decay are influenced by Solar neutrinos. These claims from experiments focus on a deficit in decay rate as a function of neutrino flux in two ways: firstly, an annual modulation in the decay rate is claimed due to annual variation in the Solar neutrino flux on Earth~\cite{Jenkins2009}. Secondly, it is claimed that neutrinos created at the Sun's atmosphere during a Solar flare cause a short-term deficit in the radioactive decay rate~\cite{Jenkins2008}. In this article, we report limits on the size of the latter effect. \\ 
The authors of~\cite{Jenkins2008} observe the coincidence of a peak in X-ray flux of $3.4\cdot10^{-4}$ W/m$^2$ (classified as X3.4) as measured by the Geostationary Operational Environmental satellite (GOES)~\cite{GOES} with a short-term decrease in the radioactive decay rate of a $\sim$1 $\mu$Ci $^{54}$Mn source during December 2006. In their study, the $^{54}$Mn sample was attached to the front of 2 $\times$ 2" NaI(Tl) detector. The number of observed decays during an 84 h period encompassing the flare was reported to be lower than the average expected decays with 7$\sigma$ significance. During the same flares as used in~\cite{Jenkins2008}, a different setup monitored the decay of the isotopes $^{90}$Sr, $^{90}$Y and $^{60}$Co using a Geiger-M\"{u}ller counter and a $^{239}$Pu source using a silicon detector~\cite{Parkhomov2010}. No significant change in the radioactive decay rate was found. Yet another study measured a $^{137}$Cs source with a HPGe detector and a $^{40}$K and a $^{nat}$Th source with a NaI(Tl) detector during Solar flares of strength X5.4 and X6.9 which occurred 2011 and 2012~\cite{Bellotti2013}. The number of counts was averaged over a 24 h period and no effect was observed, allowing those authors to report a lower limit on the fractional deviation of the decay rate to a few $0.01\%$.\\ \\
Sunspots may appear under the influence of strong magnetic fields on the Sun's photosphere. During a Solar flare, magnetic field lines at sunspots reconnect~\cite{Low1988,Smith2005} and transfer the stored magnetic energy into kinetic energy of the plasma. The heated plasma yields a strong X-ray emission and is capable of accelerating protons and electrons up to hundreds of MeVs. The typical timescale for a Solar flare is in the order of minutes~\cite{STCE2016}. Solar flares are assumed to increase the Solar neutrino flux in two ways~\cite{Bahcall1988,Fargion04}: a short-timescale increase in neutrino flux, directly created during the Solar flare, and a delayed increase in flux due to the interaction of accelerated flare particles with the Earth's atmosphere. However, neutrino flux experiments have not yet detected a correlation between Solar flares and an increase in neutrino flux~\cite{Hirata1988,Wasseige2014,Aharmim2014}.  \\ \\
On September 6th 2017, 12:02 UTC, the GOES-13 satellite observed a peak of $9.3\cdot10^{-4}$W/m$^2$ in X-ray flux~\cite{GOES,NOAA}. This flare was classified as an X9.3 flare. Four days later, on September 10th, 16:06 UTC, a second X8.2 flare was observed~\cite{GOES,NOAA}. The X-ray flux from both these flares was greater than that of the X3.4 Solar flare of December 13 2006, as well as the flares in 2011 and 2012. \\ \\
In this work, we present radioactive decay data for the month of September 2017, measured at two different locations with identical and independent NaI(Tl) detectors. The complete detector setups are discussed in detail in~\cite{Angevaare2018}. Each of the setups contain four cylindrical (76$\times$76)mm NaI(Tl) detector pairs. Three of the four detector pairs measure the de-excitation photons after the $\beta$-decay of a radioactive isotope, while the remaining pair monitors the radioactive background. Waveforms from each event are integrated to reconstruct the deposited energy and a spectrum is produced. We use a fitting routine which considers a Gaussian absorption peak, together with a background spectrum and the Compton spectrum, obtained from a Monte Carlo simulation performed with GEANT4, to obtain the number of counts inside the full absorption peak in one-hour bins~\cite{Angevaare2018}.  \\ \\
An overview of all configurations can be found in the first three columns of table \ref{TAB:Allcorr}. We search for an effect during the X9.3 Solar flare of September 6th on the decay rate of a $\sim$0.5 kBq $^{60}$Co source, continuously measured by two independent, identical detector setups located at Universit\"{a}t Z\"{u}rich (UZH), Switzerland and at Nikhef Amsterdam, the Netherlands. We also consider two $\sim$0.8 kBq sources, $^{44}$Ti and $^{137}$Cs, measured only by the detector setup at Nikhef. For the $^{60}$Co source monitored by the UZH setup we also search during the second (X8.2) flare of September 10th. Other sources monitored by the setups at UZH, Purdue University, USA and at CBPF Rio de Janeiro, Brazil~\cite{Angevaare2018} are not used here since either data was not taken during the flares, the activity of the sources was low, or the detector saturated at high energies.
\section{Methodology}
We correlate the two Solar flares, identified by their X-ray emission with the decay rate of $^{60}$Co, $^{44}$Ti and $^{137}$Cs sources. Because there is no proposed mechanism determining the size or duration of a possible effect on the decay rate, we implement three statistical techniques. This allows us to quantify the correlation between the flares and an abrupt variation in the sample data. We define multiple search periods around the time of the Solar flare, with timescales between 1 and 109 hours. The flare periods are defined as follows: the first consists of the one-hour bin coinciding with one of the two Solar flares. From this bin, other Solar flare periods are obtained by symmetrically increasing the time per bin. The maximum period of 109 hours is chosen in such a way that it encompasses the reported 84 h by~\cite{Jenkins2008} plus one day. For every search period, we calculate the relative deviation of the measured rates (see Section \ref{SEC:reldev}). Additionally, we calculate the probability of measuring the observed number of counts with respect to the expected number of counts. The first statistical technique we implement is a global significance, to correct for simultaneous testing over multiple periods (see Section \ref{SEC:twosid}). Over the reported period of 84 h, we then set a local lower limit on the relative deviation of the decay rate. We can, with only the global significance technique, not exclude the hypothesis of~\cite{Jenkins2008}. We thus perform two additional tests specifically over the reported timescale of 84 h (see Section \ref{SEC:swonesid}). We check the normality of the data set using a Shapiro Wilk (SW) normality test~\cite{SHAPIRO1965} and test the hypothesis claimed by~\cite{Jenkins2008} using a one-sided deficit test~\cite{Pillemer91}.
\subsection{Relative deviation\label{SEC:reldev}}
To calculate a variation with respect to the expected number of counts, we first perform an exponential fit to the observed number of decays. To prevent bias, we exclude the maximum considered timescale of 109 hours while fitting the exponential decay,
\begin{equation}
A(t) = A_0 \cdot 2^{-t/t_{1/2}},
\label{Eqn:Expdec}
\end{equation}
where $t$ is the time, $A_0$ the observed count rate for a particular NaI(Tl) detector at $t=0$ (midnight on September 1st UTC) and $t_{1/2}$ the half life of the isotope.
We define a unitless relative deviation in the number of counts during the period [$t_f-\frac{T}{2}$, $t_f + \frac{T}{2}$] as
\begin{equation}
\centering
\Delta N(T) = \frac{N_{\text{obs}}(T) - \mu(T)}{\mu(T)},
\label{Eqn:Reldev}
\end{equation}
where $t_f$ is the time of the Solar flare event, $T$ the timescale, $\mu(T) =\int_{t_f-\frac{T}{2}}^{t_f+\frac{T}{2}} A(t)dt$ and $N_{\text{obs}}(T)$ is the summed number of observed decays in range of [$t_f-\frac{T}{2}$, $t_f + \frac{T}{2}$]. This quantity is shown in figure \ref{FIG:Limits_Reldev} (black dots) as function of the timescale $T \in \{1,109\}$.

\subsection{Two-sided deviation limit\label{SEC:twosid}}
We perform a two-sided test~\cite{Pillemer91} for variations in the number of observed counts $N_{\text{obs}}$ with respect to the number of expected counts $\mu_i$ for each flare period and calculate the Poisson probability of finding a more extreme number than $N_{\text{obs}}$. The local test statistic is defined as 
\begin{equation}
q = 
\begin{cases}
2F(N_{\text{obs}},\mu) &\mbox{if } N_{\text{obs}} < \mu\mbox{ }\\
2[1 -F(N_{\text{obs}},\mu)] & \mbox{if } N_{\text{obs}} > \mu\mbox{ }, \\
\end{cases} 
\label{Eqn:teststatistic}
\end{equation}
where $F(N_{\text{obs}},\mu)$ is the cumulative distribution function (CDF) of the Poisson distribution which is approximated by the CDF of a Normal distribution $\Phi$ for 
\begin{equation*}
\stackunder{\text{lim}}{\scriptstyle N_{\text{obs}} \rightarrow \infty} \left(  F(N_{\text{obs}},\mu)\right) \sim \Phi(\mu,\sigma=\sqrt{\mu}).
\end{equation*} We correct for the Look-Elsewhere Effect~\cite{Lyons2008} caused by performing tests on multiple periods at the same time. To find the global significance, we define a new test statistic as 
\begin{equation}
\setstackgap{S}{2pt}
\widetilde{q} = \stackunder{\stackon{\text{min}}{\scriptstyle N}}{\scriptstyle i}(q_i),
\label{Eqn:teststatistic_mc}
\end{equation}
where $q_i$ are the local test statistics for each search period between 1 and 109 hours. We run 30,000 toy Monte Carlo (MC) simulations, which draw points from a Poisson distribution following the null hypothesis, i.e. exponential decay (described by equation \ref{Eqn:Expdec}). This MC runs over a total of 30,000 trials, chosen in order to get at least 2$\sigma$ confidence. Subsequently, we obtain the new test statistic (equation \ref{Eqn:teststatistic_mc}) over the same flare timescales between 1 and 109 hours. As a result we obtain the cumulative distribution function (CDF) of $\widetilde{q}$ under the null hypothesis, and thus translate $\widetilde{q}$ to a global p-value. Using this new global p-value, we calculate an extreme number of observed counts ($N_{obs\rightarrow extreme}$) at a particular significance level, given the expected number of counts $\mu$. \\ \\
The $1\sigma$ (blue) and $2\sigma$ (green) bands in figure \ref{FIG:Limits_Reldev} are the confidence intervals (CI) found by transforming $N_{obs\rightarrow extreme}$ into a relative deviation with equation (\ref{Eqn:Reldev}). If all points are within the global CI, we conclude that over every tested timescale our data follows exponential decay within the chosen significance level, as defined by equation (\ref{Eqn:Expdec}). \\ \\ 
We are able to set a limit with 2$\sigma$ confidence on the fractional deviation in decay rate by referencing the local test statistic. This is done by solving equation (\ref{Eqn:teststatistic}) for $q\approx0.05$, given $\mu$ is equal to the observed relative deviation over a particular period.
\subsection{Shapiro Wilk and one-sided deficit test\label{SEC:swonesid}}
To test the normality of the data set, a Shapiro Wilk test~\cite{SHAPIRO1965} is performed on the normalized residuals of a 84 h period around the Solar flare. It has been shown through simulations that, for large samples, the Shapiro Wilk test is most powerful in correctly rejecting the hypothesis for normally distributed samples~\cite{Razali2011}. If the tested sample is consistent with a normal distribution, we conclude that any observed fluctuations are consistent within bounds expected from statistical uncertainty.  \\ \\
Finally we test the hypothesis claimed by~\cite{Jenkins2008} that a Solar flare causes a decrease in rate, by performing a one-sided test on the decay rate over a period of 84 h. In this case~\cite{Pillemer91}, we test for the chance of finding a number of observed counts more extreme than the observed number of counts under the null hypothesis.
\section{Results}
\begin{figure}[t!]
	\centering
	\includegraphics[width=0.99\textwidth]{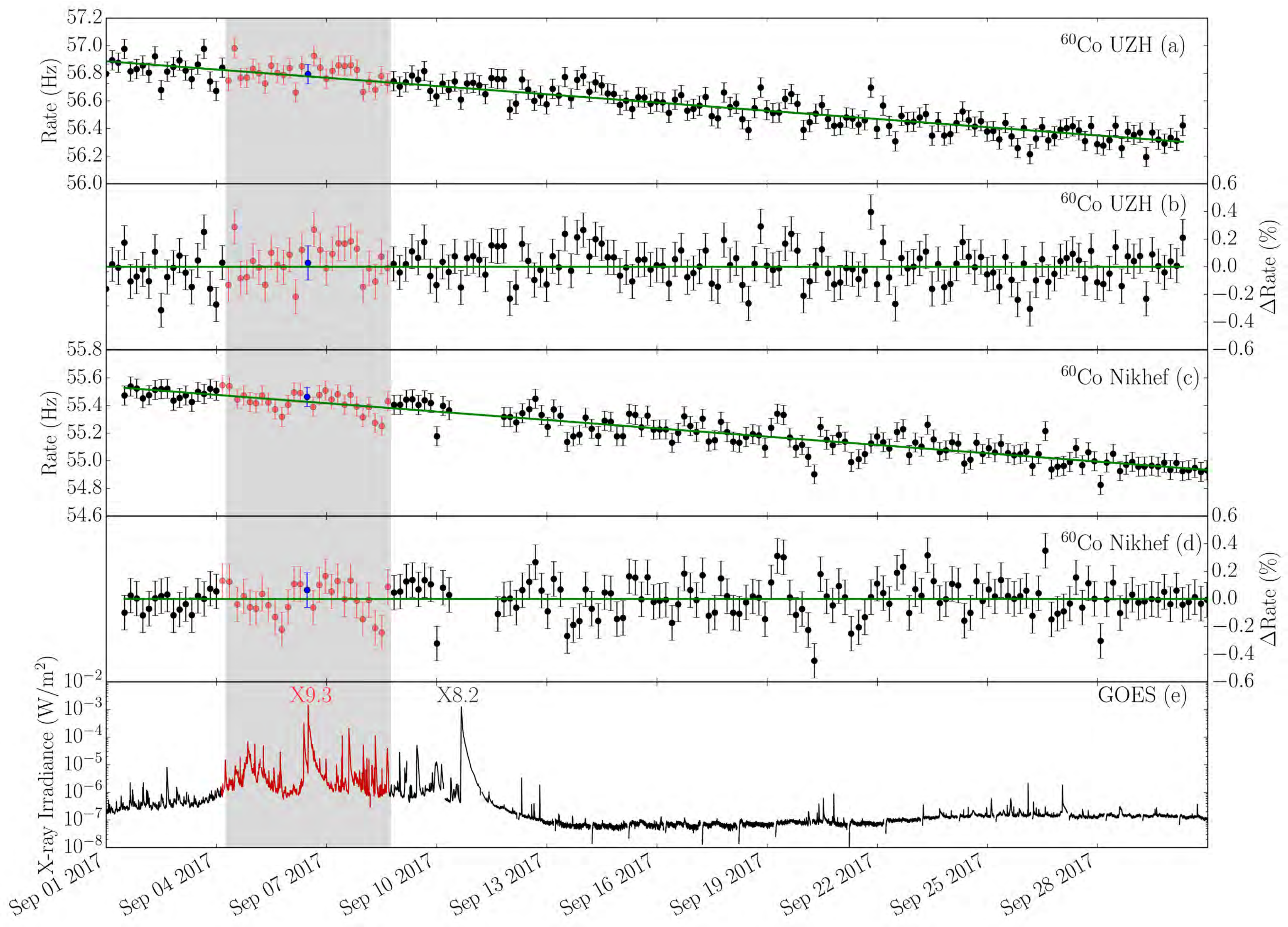}
	\captionsetup{}
	\caption{The measured count rate for $^{60}$Co decays, presented in 4 hour bins, measured by the NaI(Tl) detector setup at UZH (a) and Nikhef (c). The X-ray flux is also shown (e), as measured by the GOES satellite~\cite{GOES} during September 2017. The X9.3 Solar flare of September 6th 2017 is coincident with the blue bin. The data points of the 109 hour period around the flare (red) are excluded from the exponential fit (green) which is used to determine the residuals (c and d for the UZH and Nikhef NaI(Tl) setups, respectively). The same procedure is followed for the analysis of X8.2 Solar flare.}
	\label{FIG:Rate_Co60}
	
\end{figure}

\begin{figure}[b!]
	\centering
	\captionsetup{}
	\includegraphics[width=\textwidth]{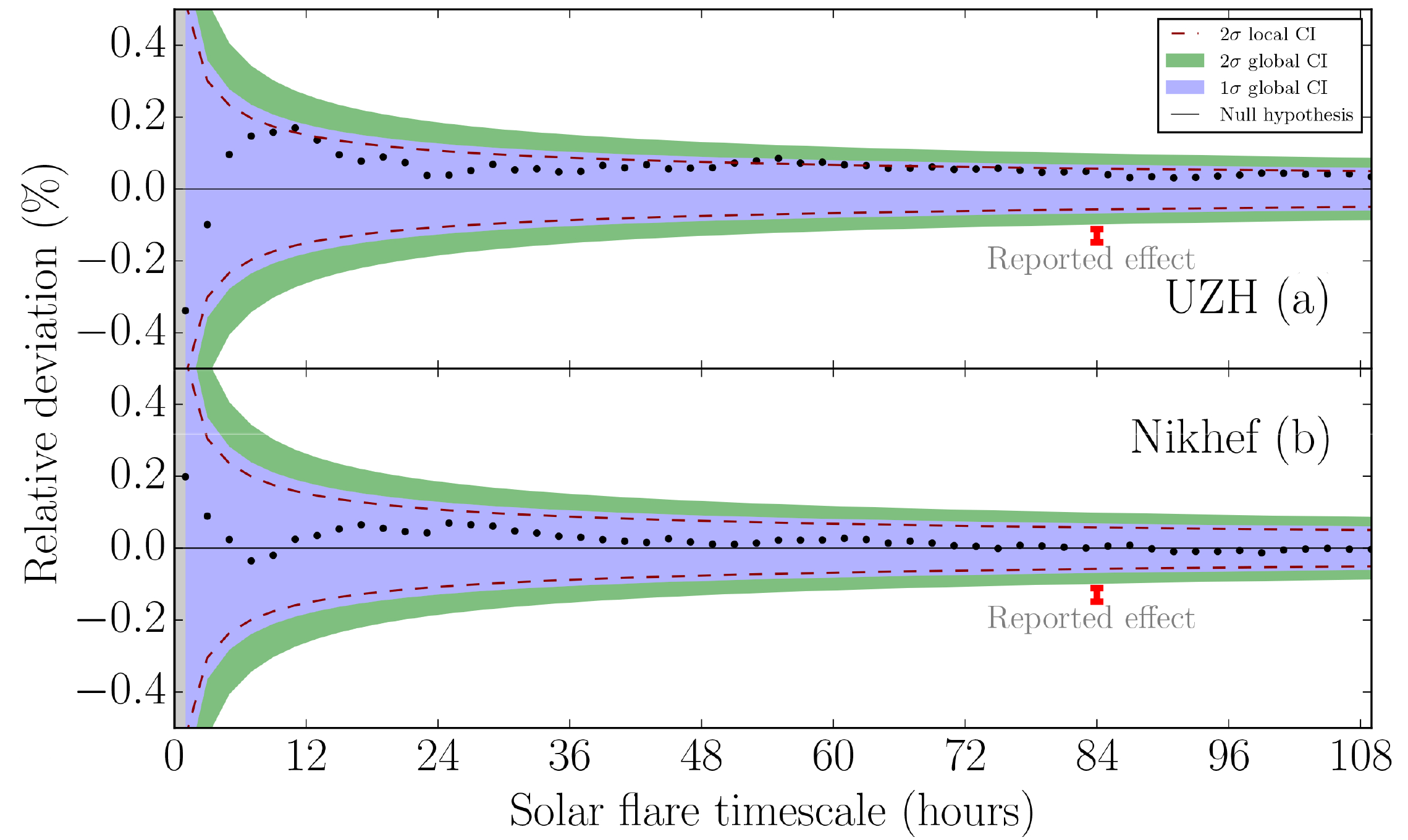}
	\caption{The result of the two-sided test for the decay of $^{60}$Co. The black line corresponds to the null hypothesis of pure exponential decay. Every data point corresponds to a relative deviation with respect to the expectation, calculated by equation (\ref{Eqn:Reldev}), over timescales around the Solar flare ranging from 1 to 109 hours for the UZH (a) and Nikhef data (b). The global 1$\sigma$ (blue), global 2$\sigma$ (green) and local 2$\sigma$ (red dashed) significance intervals are also included. The red data point denotes the effect reported by~\cite{Jenkins2008}.}
	\label{FIG:Limits_Reldev}
\end{figure}

Figure \ref{FIG:Rate_Co60} shows the measured decay rate of $^{60}$Co over the month of September 2017 at UZH and at Nikhef, along with the X-ray flux including the X9.3 and X8.2 Solar flares, as measured by the GOES-13 satellite~\cite{GOES}. The detector setup at Nikhef did not collect any data due to a data storage error between September 10th, 11:37 UTC and September 11th, 10:59 UTC, which encompasses the second X8.2 Solar flare event. 
All data shown is binned in four-hour periods. The normalized residuals are shown, calculated by

\begin{equation}
\Delta\text{Rate} = \frac{N_{\text{obs}}-\mu}{\mu},
\end{equation} which is equation (\ref{Eqn:Reldev}) evaluated individually for every bin. There is no clear effect observed in the measured decay, coinciding with the flare.\\
Figure \ref{FIG:Limits_Reldev} shows the relative deviation from a pure exponential for the decay rate of $^{60}$Co, measured by the UZH and Nikhef detector setups for timescales between 1 and 109 hours around the X9.3 flare, including the 1$\sigma$ (blue) and 2$\sigma$ (green) global confidence intervals. Note that, following equation (\ref{Eqn:Reldev}), all points are correlated. We conclude that all deviations are inside the bounds expected from statistical uncertainty. Within 1 hour around the Solar flare, the relative deviation observed in both setups is \textless 0.4\% and well within the global 2$\sigma$ confidence interval. The UZH setup (a) observes an insignificant deficit, while the Nikhef setup (b) observes an insignificant excess. For both setups all points are within the global 2$\sigma$ confidence interval. \\ \\
At Nikhef, a $^{137}$Cs and a $^{44}$Ti source, both of $\sim$0.8 kBq activity, were also monitored during September 2017. Following the same analysis, we find no significant deviations in those decay rates during the X9.3 flare. We follow the same analysis for the $^{60}$Co decay rate measured at UZH during the second X8.2 Solar flare and found no significant deviations in the decay rate. We summarize our results in table \ref{TAB:Allcorr}. We therefore conclude that we observe no significant correlation between the decay rate and the Solar flares, and that all measured fluctuations in this work paper are consistent with bounds from statistical uncertainties. \\ \\
From the continuously monitored slow control parameters we find that the variation of the temperature, PMT high voltage, magnetic field strength and radon activity are within operating range as defined by table 4 in~\cite{Angevaare2018}. We therefore conclude that systematic influences due to slow control parameters remain \textless $ 2\cdot10^{-5}$. The pressure and humidity were also monitored and remained stable during the full measurement period with 0.8\% and 17\% at the 68\% CL for both setups, respectively.\\ \\
Table \ref{TAB:Allcorr} summarizes the results of this study for each isotope. The lower limits, the p-values of the SW test and p-values of the one-sided deficit test are presented for every monitored isotope over a period of 84 h. The SW test p-values (column 5) are \textgreater0.005. The lowest p-value of 0.005 is found due to an outlier within the tested period. Despite the outlier, the p-values from the one-sided deficit test (column 6) are \textgreater0.4. From these p-values we conclude that for a period of 84 h around the Solar flare, the distribution of the measured decay rate is consistent with a normal distribution and that we find no significant deficit with respect to the expectation.\\ \\
\begin{table*}[t]
	\caption{Results of correlation of the decay rates of $^{60}$Co, $^{137}$Cs and $^{44}$Ti and both Solar flares, measured by two different setups. The SW test, 2$\sigma$ limit and the one-sided test were all performed on a period of 84 h.}
	\centering
	\begin{tabular}{ c  c  c  c  c  c  c }
		\hline
		\small\textbf{Isotope} & \small\textbf{Flare classi-}&\small\textbf{Setup} & \small\textbf{2$\sigma$ Lower Limit } & \small\textbf{SW test}&\small\textbf{One-sided test} \\
		& \small\textbf{fication}& & \small(\%) & \small p-value & \small p-value\\
		\hline 
		$^{60}$Co & X9.3 & Zurich &$-0.019$& 0.2  & 0.97 \\
		$^{60}$Co & X9.3 & Nikhef & $-0.062$& 0.5& 0.54  \\ \hdashline
		$^{137}$Cs & X9.3 & Nikhef & $-0.033$&0.08 & 0.40 \\
		$^{44}$Ti & X9.3 & Nikhef & $-0.009$&0.005& 0.86  \\ \hdashline
		$^{60}$Co & X8.2 & Zurich &  $-0.046$ & 0.5  & 0.71 \\
	\end{tabular}
	\label{TAB:Allcorr}
\end{table*}Before we set a limit using the local test statistic (equation \ref{Eqn:teststatistic}) and exclude the claimed hypothesis by~\cite{Jenkins2008}, we correct the distribution of the test statistic $q$ for fitting over a finite number of data points. Without this correction, one would assume $q$ to be distributed as $q\sim\mathcal{N}(\mu,\sqrt{\mu})$ and therefore underestimate the width of this distribution. The correction is calculated by performing a fit on MC generated datasets, assuming the null hypothesis, (see Section \ref{SEC:twosid}) with the same number of data points as in the analysis of our measured data (presented in figure \ref{FIG:Rate_Co60}). The weakest, and thus most conservative, exclusion and limit are found for the decay of $^{60}$Co, measured by the setup at Nikhef (figure \ref{FIG:Limits_Reldev} (b)) during the X9.3 flare. We exclude the hypothesis of~\cite{Jenkins2008} with 4.7$\sigma$ confidence and set a lower limit on the deficit in decay rate to 0.062\% with 2$\sigma$ confidence. The lower limits of the other measured isotopes are listed in column four in table \ref{TAB:Allcorr}. 
\section{Conclusion}
We find no evidence for a deficit or excess in the decay rate of three measured radioactive isotopes during periods of Solar flare activity. A decrease in decay rate similar to the one reported by~\cite{Jenkins2008} is not found in either of our setups for $^{60}$Co, $^{137}$Cs and $^{44}$Ti decays during two different Solar flares. We exclude the hypothesis reported by~\cite{Jenkins2008} at 4.7$\sigma$ confidence and set a lower limit with $2\sigma$ confidence on the deficit in decay rate for a period of 84 h level at 0.062$\%$, which is $\sim2$ times smaller than the reported effect of $\sim0.1\%$~\cite{Jenkins2008}. These results are consistent with findings of~\cite{Bellotti2013}, who also do not observe a fractional effect larger than a few $0.01\%$ in the decay rate over a tested period of 24 hours. 
\section*{Acknowledgements}
We are grateful for the support of the Purdue Research Foundation, the Netherlands Organization for Scientific Research (NWO), the University of Zurich, the Swiss National Science Foundation under Grant Nos. 200020-162501 and the Foundation for Research Support of the State of Rio de Janeiro (FAPERJ).
\section*{References}
\bibliographystyle{elsarticle-num}
\providecommand{\href}[2]{#2}\begingroup\raggedright\endgroup
\end{document}